\newcommand{\msun}{\mbox{${\rm M}_\odot$}}

\newcommand{\nbody}{\mbox{$N$-body}}
\newcommand{\Wo}{\mbox{$W_0$}}

\newcommand{\kira}{\mbox{\sf kira}}
\newcommand{\SeBa}{\mbox{\sf SeBa}}

\newcommand{\tlast}{\mbox{$t_{\rm last}$}}

\newcommand{\ncoll}{\mbox{${\cal N}_{\rm coll}$}}
\newcommand{\rcoll}{\mbox{${\cal R}_{\rm coll}$}}
\newcommand{\fcoll}{\mbox{${\rm f}_{\rm coll}$}}
\newcommand{\nrun}{\mbox{${\cal N}_{\rm run}$}}
\newcommand{\rgc}{\mbox{$r_{\rm GC}$}}
\newcommand{\rtide}{\mbox{$r_{\rm tide}$}}
\newcommand{\trlx}{\mbox{$t_{\rm rlx}$}}
\newcommand{\tcrss}{\mbox{$t_{\rm crss}$}}

\newcommand{\rhm}{\mbox{$r_{\rm hm}$}}

\documentstyle[12pt,./pasp,./psfig]{article}
\def\unit#1{{\mbox{[{\rm #1}]}}}
\def\apgt{\ {\raise-.5ex\hbox{$\buildrel>\over\sim$}}\ }
\def\aplt{\ {\raise-.5ex\hbox{$\buildrel<\over\sim$}}\ }
\begin{document}

\title{Runaway collisions in star clusters}

\bigskip
\bigskip

\author{Simon F.\ Portegies Zwart\altaffilmark{1},
	Junichiro Makino\altaffilmark{2},
	Stephen L.\ W.\ McMillan\altaffilmark{3},
    	Piet Hut\altaffilmark{4}
}

\altaffiltext{1}{Massachusetts Institute of Technology, Cambridge, MA
		 02139, USA, Hubble Fellow}
\altaffiltext{2}{Department of Astronomy, University of Tokyo, 7-3-1 Hongo,
       Bunkyo-ku,Tokyo 113-0033, Japan}
\altaffiltext{3}{Department of Physics,
		 Drexel University,
                 Philadelphia, PA 19104, USA}
\altaffiltext{4}{Institute for Advanced Study,
		 Princeton, NJ 08540, USA}

\bigskip
\bigskip

%\slugcomment{Simon Portegies Zwart is a Hubble Fellow}

\begin{abstract}
We study the occurrence of physical collisions between stars in young
and compact star cluster.  The calculations are performed on the
GRAPE-4 with the starlab software environment which include the
dynamical evolution and the nuclear evolution of all stars and
binaries.  The selection of the initial conditions is based on
existing and well observed star clusters, such as R136 in the 30
Doradus region in the Large Magellanic Cloud and the Arches and
Quintuplet star clusters in the vicinity of the Galactic center.
Collisions between stars occurred rather frequently in our models.  At
any time a single star dominates the collision history of the system.
The collision rate of this runaway merger scales with the initial
relaxation time of the cluster and is independent on other cluster
parameters, such as the initial mass function or the initial density
profile of the cluster.  Subsequent encounters result in a steady grow
in mass of the coagulating star, until it escapes or explodes in a
supernova.  The collision rate in these models is about $2.2\times
10^{-4}$ collisions per star per Myr for a cluster with an initial
relaxation time of 1\,Myr.

\end{abstract}

\section{Introduction}

The central regions of young star clusters, globular clusters and
galaxies have a high stellar density and close encounters between
stars are thought to occur on a regular basis.  In some cases stars
may even collide and form new objects.  However, proof for the occurrence of
such collisions has never been found explicitly in observations or in
theoretical models.  Direct evidence for collisions could be provided
by the observation of a blue straggler with a mass exceeding three
times the mass of the clusters' turn off, by discovery of a blue
straggler in an eccentric orbit around another main-sequence star or
by catching a collision in the act.

There are quite a few indications that collisions play a major role in
the evolution of star clusters and their constituents.  For example:
the presence of blue stragglers and millisecond pulsars in globular
clusters indicates that stellar collisions may play a role, even
though alternative --non-collisional-- scenarios have also been
proposed (Leonard 1989).\nocite{1989AJ.....98..217L}

Collision rates in star clusters have always been estimated using
cross section arguments. Pioneering work in this field has been
performed by Hills (1975) and Lightman \& Shapiro
(1977).\nocite{1975AJ.....80..809H}\nocite{1977ApJ...211..244L}
Their general conclusions are:
\begin{itemize}
\item[1)] In order to get a runaway growth of a single star one requires 
	  a star cluster which contains typically more than $\sim 10^7$ stars. 
\item[2)] A massive star formed out of multiple collisions will always be
	  accompanied by many objects which experienced only
	  one or two collisions.
\item[3)] Stellar collisions are so rare that they do not affect the
	  evolution of the star cluster.
\end{itemize}

We show that these results are based on insufficient detail in the
models and that the contrary has to be concluded:
\begin{itemize}
\item[ad 1)] Runaway growth is a natural phenomena which occurs even
in rather small cluster.
\item[ad 2)] The massive product of many collisions will therefore be
well separated in mass from the other objects.
\item[ad 3)] Even in star clusters where physical collisions are
relatively rare the evolution of the star cluster is altered
significantly.
\end{itemize}

We will address these issues by the direct integration of the orbits
of all the stars in such clusters, while accounting for the tidal
field of the Galaxy and the internal evolution of the stars and
binaries.  Our model calculations are performed using the starlab
software environment with up to 32k stars. The calculations are sped
up with the special purpose computer GRAPE-4.  The initial conditions
are taken to represent observed compact star clusters such as R136
(the compact star cluster in the 30 Doradus region), the Arches
cluster and the Quintuplet system.

\subsection{The model}

To model a dense star cluster we use a hydride computer program that
consists of two independent parts. An \nbody\ model integrates the
equations of motion of all stars and at the same time the other part
computes the evolution of the stars.  The back coupling between the
stellar evolution and the stellar dynamics is taken into account in a
self consistent --object oriented-- fashion.

All computations are performed using the starlab toolset (Portegies
Zwart et al.\, 2000b, see also {\tt
http://www.manybody.org}).\nocite{2000astro.ph.5248PZ} Starlab
consists of the \nbody\, integrator \kira\, and the stellar and binary
evolution program \SeBa. The calculations are accelerated using the
special purpose computer GRAPE-4 (Makino et al.\
1997).\nocite{1997ApJ...480..432M}

%%%%%%%%%%%%%%%%%%%%%%%%%%%%
\section{Initial conditions}

Selection for the initial conditions in our calculations were
guided by the well studied Arches and the Quintuplet clusters and 
the central star cluster R136 in the 30 Doradus region of the Large
Machelanic cloud.  A compilation of the characteristics of these
clusters is presented in Tab.\,1.  The Arches and Quintuplet cluster
are very close to the Galactic center where R136 is completely
isolated from the tidal perturbation of the Galaxy.

\begin{table*}[ht]
\caption[]{Observed  parameters  for  R\,136, the  Arches and the Quintuplet
systems.  Columns list cluster name, reference, age, mass, projected
distance to the Galactic center, tidal radius ($\rtide$), and half
mass radius ($\rhm$).  The final column presents an estimate of the
density within the half mass radius.  }
\begin{flushleft}
\begin{tabular}{ll|rrrrrr} \hline
Name 	  &ref& Age  &   M     & \rgc & \rtide & \rhm   
				& $\log \rho_{\rm core}$ \\ 
          &&[Myr]& [$10^3$\,\msun] & \multicolumn{3}{c}{------ [pc] ------}   &
			 [\msun/pc$^3$] \\ \hline
R\,136    &a& 2--4 & 21--79   & 50k  &$\apgt20$&$\sim 0.5$& 4.6--5.2 \\
Arches    &b& 1--2 & 12--50   & 30   & 1       &  0.2     & 5.6--6.2 \\ 
Quintuplet&c& 3--5 & 10--16   & 50   & 1       &$\sim 0.5$& 4.3--4.5 \\
\hline
\end{tabular} \\
\smallskip
References:
a) Brandl et al.\, (1996);\nocite{1996ApJ...466..254B} 
   Campbell et al.\, (1992);\nocite{1992AJ....104.1721C}
   Massey \& Hunter (1998).\nocite{1998ApJ...493..180M}
b) Figer et al.\, (1999);\nocite{1999ApJ...525..750F}
c) Glass, Catchpole \& Whitelock (1987);\nocite{1987MNRAS.227..373G} 
   Figer, Mclean \& Morris (1999).\nocite{1999ApJ...514..202F} 
\end{flushleft}
\label{Tab:observed} 
\end{table*}

We performed a total of 44 $N$-body calculations over a wide range of
initial conditions. The number of stars was varied from 1k (1024) to
32k. Initial density profiles and velocity dispersion for the models
are taken from Heggie-Ramamani models (Heggie \& Ramamani
1995)\nocite{1995MNRAS.272..317H} with \Wo\, ranging from 1 to 7.  At
birth, the clusters are assumed to perfectly fill the zero velocity
surface in the tidal field of the Galaxy. In most cases we selected
the initial mass function between 0.1\,\msun\ and 100\,\msun\,
suggested for the Solar neighborhood by Scalo (1986),\nocite{scalo86}
but several calculations are performed with power law initial mass
functions with an exponent of -2 and -2.35 (Salpeter).  The
specification of the remaining parameters, such as the strength of the
Galactic tidal field, the virial radius and the tidal radius, are
discussed in a forthcoming paper. Here we only specify the initial
relaxation time at the half mass radius of our models as this happens
to be the fundamental parameter with which our results can be
understood (see sect.\,\S\,\ref{Sect:results}).  An overview of the
initial conditions for the computed models is summarized in
Tab.\,\ref{Tab:models}.  A detailed discussion about several of the
presented calculations by Portegies Zwart et
al. (2001).

\begin{table}[htbp!]
\caption[]{Overview of the calculations. 
The first column gives the name of the model as used in previous
publications (names RxWx and KMLx are from Portegies Zwart et al. 2001
[see also Portegies Zwart et al. 2000a]; and the other models are
described in detail by Portegies Zwart et al 1999).\nocite{pzmmh99}
The following columns give the number of stars (in units of 1024), the
initial mass function (Scalo 1986, or a power law with slope as
indicated), the initial King \Wo, the initial relaxation time (in Myr)
and the number of runs performed with these initial conditions
(\nrun). The last three columns give the average number of collisions
in these calculations, the moment the last collision occurred and the
collision rate per Myr per star (see Eq.\,\ref{Eq:fcoll}.  The models
indicated with $\star$ have been computed without a Galactic tidal
field (see Portegies Zwart 1999).  }
\begin{flushleft}
\begin{tabular}{llclrlrrr} \hline
model&
$\langle N\rangle$&
IMF&
$\langle \Wo\rangle$&
$\langle \trlx\rangle$&
\nrun&
$\langle \ncoll \rangle$&
$\langle \tlast \rangle$&
\fcoll \\
\hline
%            N  TF   IMF Wo Trlx  Nrun  Ncoll    Tlast  Fcoll 
R34W7  &   12k &  Scalo&   7& 0.4&    2&  16.&    10.4& -3.70  \\
KML112 &    4k &  -2   &   7& 0.5&    2&  4.0&    1.9 & -3.00  \\
KML101 &    4k &  -2   &   4& 1.4&    2&  2.0&    1.0 & -3.60 \\  
KML142 &    6k &  -2.35&   4& 1.9&    1&  1.0&    2.2 & -4.13  \\
KML111 &    4k &  -2   &   1& 2.3&    2&  0.5&    6.7 & -4.95 \\
R90W7  &   12k &  Scalo&   7& 2.8&    1&  13.&    10.0& -4.44  \\
R34W4  &   12k &  Scalo&   4& 3.2&    3&  6.3&    30.0& -4.26  \\
KML144 &   14k &  -2.35&   4& 3.9&    1&  2.0&    2.4 & -4.18  \\
R150W7 &   12k &  Scalo&   7& 4.5&    2&  10.&    21.3& -4.52  \\
6k6X5$^\star$  		    
       &    6k &  Scalo&   6& 5.0&    1&  21.&    47.9& -3.47  \\
R34W1  &   12k &  Scalo&   1& 5.5&    3&  4.7&    29.1& -4.28  \\
R90W4  &   12k &  Scalo&   4& 8.1&    5&  5.8&    10.0& -4.47  \\
Nk6X10$^\star$ 		    
       &    9k &  Scalo&   6&10.0&    8&  10.&    18.0& -3.94  \\
R150W4 &   12k &  Scalo&   4&13.0&    4&  8.5&    7.3 & -4.70  \\
R90W1  &   12k &  Scalo&   1&14.6&    1&  7.0&    9.8 & -4.57  \\
6k6X20$^\star$ 		    
       &    6k &  Scalo&   6&20.0&    2&  4.0&    95.4& -4.19  \\
R150W1 &   12k &  Scalo&   1&23.6&    2&  3.0&    2.1 & -4.87  \\
R300W4 &   12k &  Scalo&   4&55.6&    1&  1.0&    10.0& -5.77  \\
R34W1  &   32k &  Scalo&   1&58.1&    1&  4.0&    35.8& -5.89  \\
\hline     		       
\end{tabular}
\end{flushleft}
\label{Tab:models} \end{table}

The \nbody\, system is fully determined once the tidal field, the
initial mass function, the number of stars and the relaxation time of
the cluster are selected.  The total mass of the stellar system
determines the unit of mass in the \nbody\, system, the tidal radius
\rtide\, sets the distance unit and the velocity dispersion together
with the size of the stellar system sets the time scale (see Heggie \&
Mathieu 1986).\nocite{HM1986} The evolution of the cluster is
subsequently followed using the direct
\nbody\, integration including stellar and binary evolution and the
tidal field of the Galaxy (see Portegies Zwart et al.\,
2000b).\nocite{2000astro.ph..8490P} For economic reasons not all stars
are kept in the \nbody\, system, but stars are removed when they are
3\rtide\, from the center of the star cluster.

The evolution of our star clusters is driven by two-body relaxation,
by stellar mass loss and by the external tidal field of the parent
Galaxy. The initial relaxation time, however, appears to be the most
fundamental parameter for the range of parameters we study here.  The
two-body relaxation time:
\begin{equation}
	t_{\rm rlx} \propto {N \over \ln (\gamma N)} \tcrss.
\label{Eq:trlx}\end{equation}
Here $N$ is the number of stars and $\gamma$ is a scaling factor,
introduced to model the effects of the cut-off in the long range
Coulomb logarithm (see Giertz \& Heggie 1994;
1996).\nocite{1996MNRAS.279.1037G}\nocite{1994MNRAS.268..257G} Here
\tcrss\ is the half-mass crossing time of the cluster is
\begin{equation}
\tcrss \simeq 57 \left( {[\msun] \over M} \right)^{1/2}
                 \left( {\rhm \over [{\rm pc}]}\right)^{3/2} \, \unit{Myr}.
\label{Eq:thc}\end{equation}
Here $r_{\rm hm}$ is its half mass radius and $M$ is the mass of the cluster.  

\subsection{Scaling the collision cross section}\label{Sect:collscale}
The rate at which stars in a cluster experience collisions can be
estimated via
\begin{equation}
 	n_{\rm coll} \propto n_c \sigma v.
\label{Eq:rate}\end{equation}
Here $n_c$ is the number density of the stars in the core, $\sigma$ is
the collision cross section (for approach within some distance $d$),
and $v$ is the velocity dispersion of the cluster stars.  

The central density and the collision cross section are given by the
following proportionalities:
\begin{eqnarray}
	n_c 	&\propto& 	{N \over r_c^3}, \nonumber	\\
	\sigma 	&\propto& 	d^2 + {d \over v^2}. 
\label{Eq:crossection}\end{eqnarray}
We neglect the $d^2$ term in the cross section. In the case of
constant crossing time, $Nr^3 = {\rm constant}$, we may write $v
\propto N^{1/3}$.  The number of collisions is then computed by
substitution of Eq.\,(\ref{Eq:crossection}) into Eq.\,(\ref{Eq:rate})
to obtain $n_{\rm coll} \simeq {d N^{-1/3}}$ resulting in a
collision rate per star per relaxation time of:
\begin{equation}
 	\rcoll \simeq {d N^{2/3} \over \trlx}. 
\label{Eq:ncoll}\end{equation}

From a theoretical perspective one expects therefore that the
collision rate scales with the number of stars and inversely with the
initial relaxation time of the stellar system. The normalization
constant, however, is hard to estimate from first principals.

\section{Results}\label{Sect:results}

Shortly after the start of our calculations, each model experiences
core collapse. This event is triggered by the more massive stars in
the cluster, which sink into the cluster core in a fraction of the
initial relaxation time; $\sim \trlx m/\langle m \rangle$.  Once in
the cluster core these massive stars dominate the dynamics by forming
binaries with other stars.  This phase of cluster evolution continues
until the massive star is either kicked out of the cluster by a strong
encounter or the star explodes in a supernova (see Portegies Zwart et
al 1999).  The period over which the massive star can dominate the
cluster dynamics depends on the initial conditions of the star
cluster. 

All our models finally dissolve in the tidal field of the Galaxy.
(Except the isolated models which are terminated after one initial
relaxation time lapsed.)

Once a massive star settles in the cluster center and forms a binary
with another star the cluster tends to becomes dominated by
collisions.  For an detailed description of this process we refer to
Portegies Zwart et al. (1999). 

\subsection{The collision rate from model calculations}

The number of collisions in each simulation ranges from 0--24,
resulting a rate of $10^{-6}$---$10^{-3}$ collision per star per Myr
(see Tab.\,\ref{Tab:models})\footnote{Collision rates are computed over
the time interval from zero age to the moment the last collision
occurs.}. These numbers are about two orders of magnitude higher than
expected from simple cross section arguments. The reason for this
discrepancy hides in the effect of mass segregation and the formation
of binaries. 

\begin{figure}[htbp!]
\hspace*{1.cm}
\psfig{figure=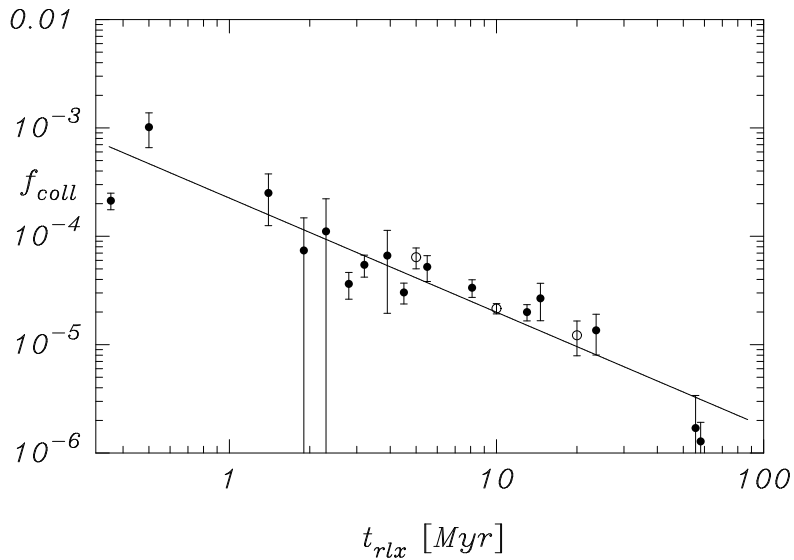,width=12.cm,angle=-90}
\caption[]{Collision rate as function of the initial relaxation time
for all models presented in Tab.\,\ref{Tab:models}.  The open circles
give the results of our systems which are isolated from the Galactic
potential (see Portegies Zwart et al 1999).  Vertical bars represent
Poissonian one-$\sigma$ errors.  The solid line is a least squares fit
to the data.  }
\label{Fig:ncoll_Trlx}
\end{figure}

The number of collisions depends on the number of stars and on the
initial relaxation time of the stellar system.  
We therefore define \fcoll\, as:
\begin{equation}
	\fcoll = {\ncoll \over N \tlast}.
\label{Eq:fcoll}\end{equation}
Here \tlast\, is the time of the last collision.  In
Fig.\,\ref{Fig:ncoll_Trlx} we give the collision rate \fcoll\, per
star per million years as a function of the initial relaxation time of
the model cluster. 

Models with the same initial conditions (Galactic tidal field,
relaxation time and mass function) are averaged.

The solid line in Fig\,\ref{Fig:ncoll_Trlx} is a least squares fit to
the results of our \nbody\, calculations, and yields
\begin{equation}
	\rcoll = 2.2 \times 10^{-4} \trlx^{-1.0} \; \; 
		       [{\rm s}^{-1}].
\label{Eq:fit}
\end{equation}

The fit (Eq.\,\ref{Eq:fit}) is excellent, which is somewhat surprising
since the models range a large area of parameter space, in initial
density profile, mass function and tidal field (or not).  Apparently
these parameters hardly affect the collision rate.  The fundamental
parameters appear to be the initial relaxation time of the stellar
system \trlx\, and the number of stars $N$.

%%%%%%%%%%%%%%%%%%%%%%%%%%%
\subsection{The collision partners}
\begin{figure}[htbp!]
\hspace*{1.cm}
\psfig{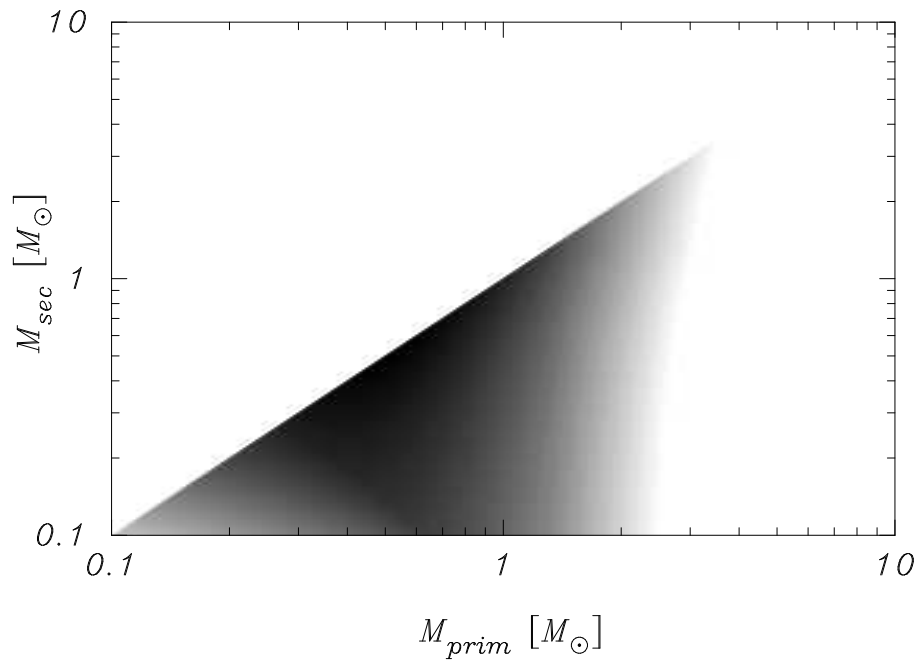}
\caption[]{
Distribution for collision candidates from derived using a
Scalo (1986) mass function between 0.1\,\msun\, and 100\,\msun\, at
zero age.  }
\label{Fig:expected_rate}
\end{figure}

From cross section arguments one would expect that the most
common collision candidates are among the stars which have the largest
collective cross section (the collective projected area of the stars
convolved with their gravitational focusing).
Fig.\,\ref{Fig:expected_rate} gives the probability distribution of
the stars which are most likely to be involved in a collision; the
most likely collision counterparts are rather low mass stars, which
are most common (see Portegies Zwart et al. 1999 for details).

\begin{figure}[htbp!]
\hspace*{1.cm}
\psfig{figure=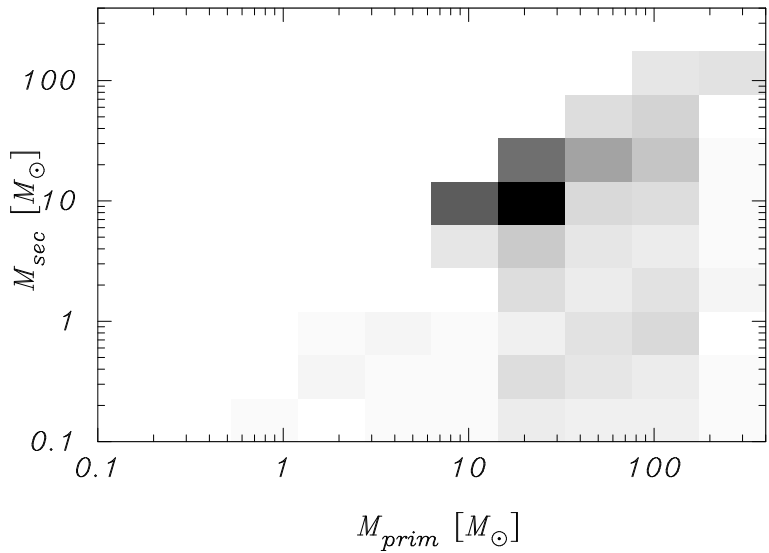,width=12.cm,angle=-90}
\caption[]{
Probability distribution for collision candidates from our model
calculations with Scalo (1986) initial mass function as in
Fig.\,\ref{Fig:expected_rate}. Darker shades indicates that more
collisions between stars with these masses occurred in our
calculations.  }
\label{Fig:model_rate}
\end{figure}

Fig.\,\ref{Fig:model_rate} gives the probability distribution for the
stars that did experience a collision in our model calculations. The
data are taken from a rather inhomogeneous sample of calculations (see
Tab.\,\ref{Tab:models}).

The most striking difference between Fig.\,\ref{Fig:expected_rate} and
Fig.\,\ref{Fig:model_rate} is the mean mass of the colliding stars.
Where Fig.\,\ref{Fig:expected_rate} only extends to 10\,\msun\,
stars and gives a highest probability for a collision between
0.7\,\msun\, stars, Fig.\,\ref{Fig:model_rate} shows that much higher
mass stars are much more likely to experience collisions. The mean mass
of the colliding stars according to the model calculations exceeds
10\,\msun.  The reasons for these discrepancies hide in the effect of
mass segregation and the formation of binaries in the core of the
stellar system (see Portegies Zwart et al.\, 1999 for details).

%%%%%%%%%%%%%%%%%%%%%%%%%%%
\subsection{Runaway growth}

The mean mass of a collision counterpart is about 10\,\msun\, and many
collisions involve the same star, the growth rate of this star is
about 10\,\msun\ per collision. Once the collision runaway explodes in
a supernova or is ejected from the stellar system its growth stops and
another star takes its place until the cluster dissolves in the tidal
field of the Galaxy. (The models which were calculated without an
external potential were stopped well before the cluster dissolved.)

\begin{figure}[htbp!]
\hspace*{1.cm}
\psfig{figure=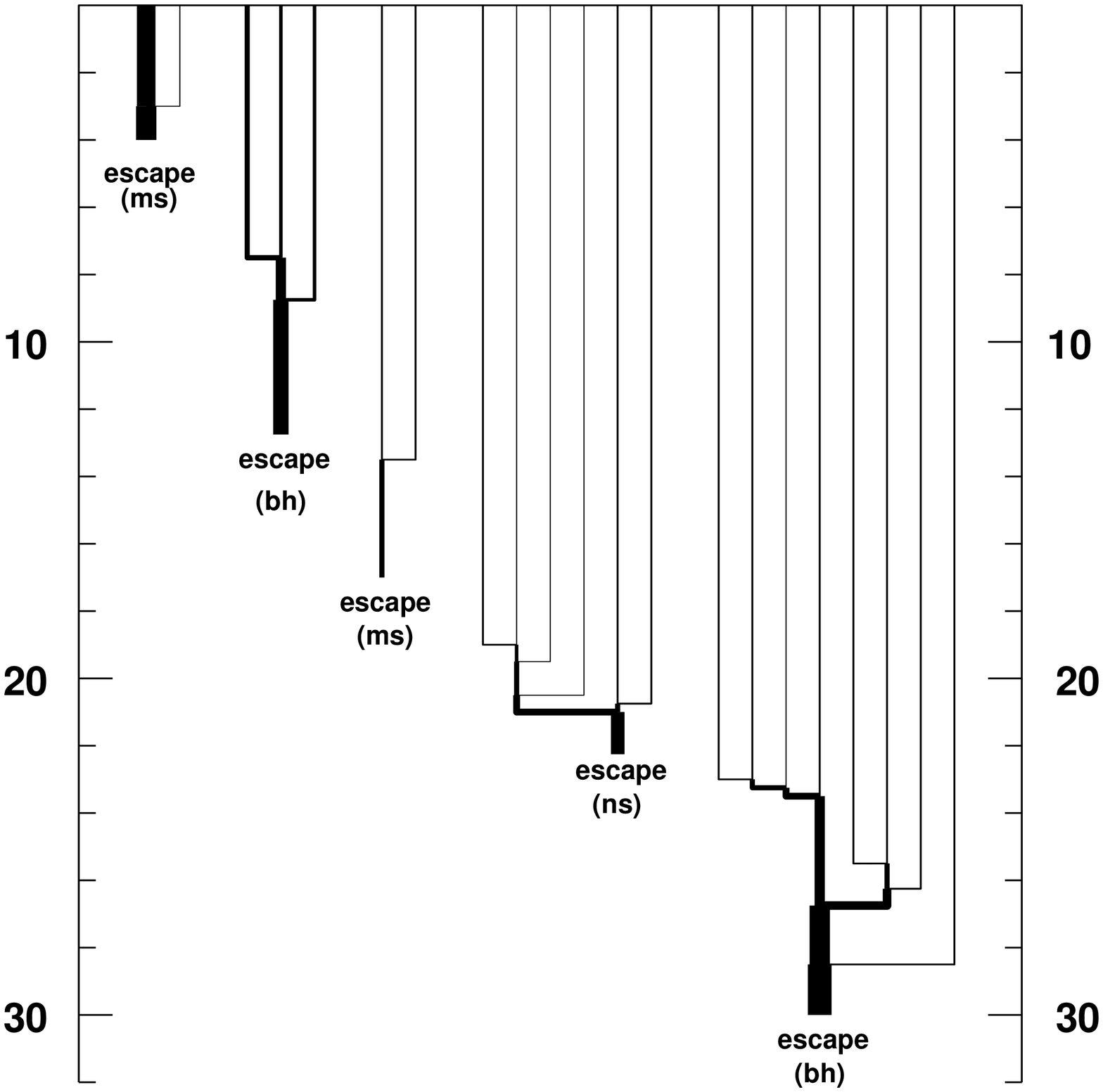,width=7.2cm,angle=-90}
\psfig{figure=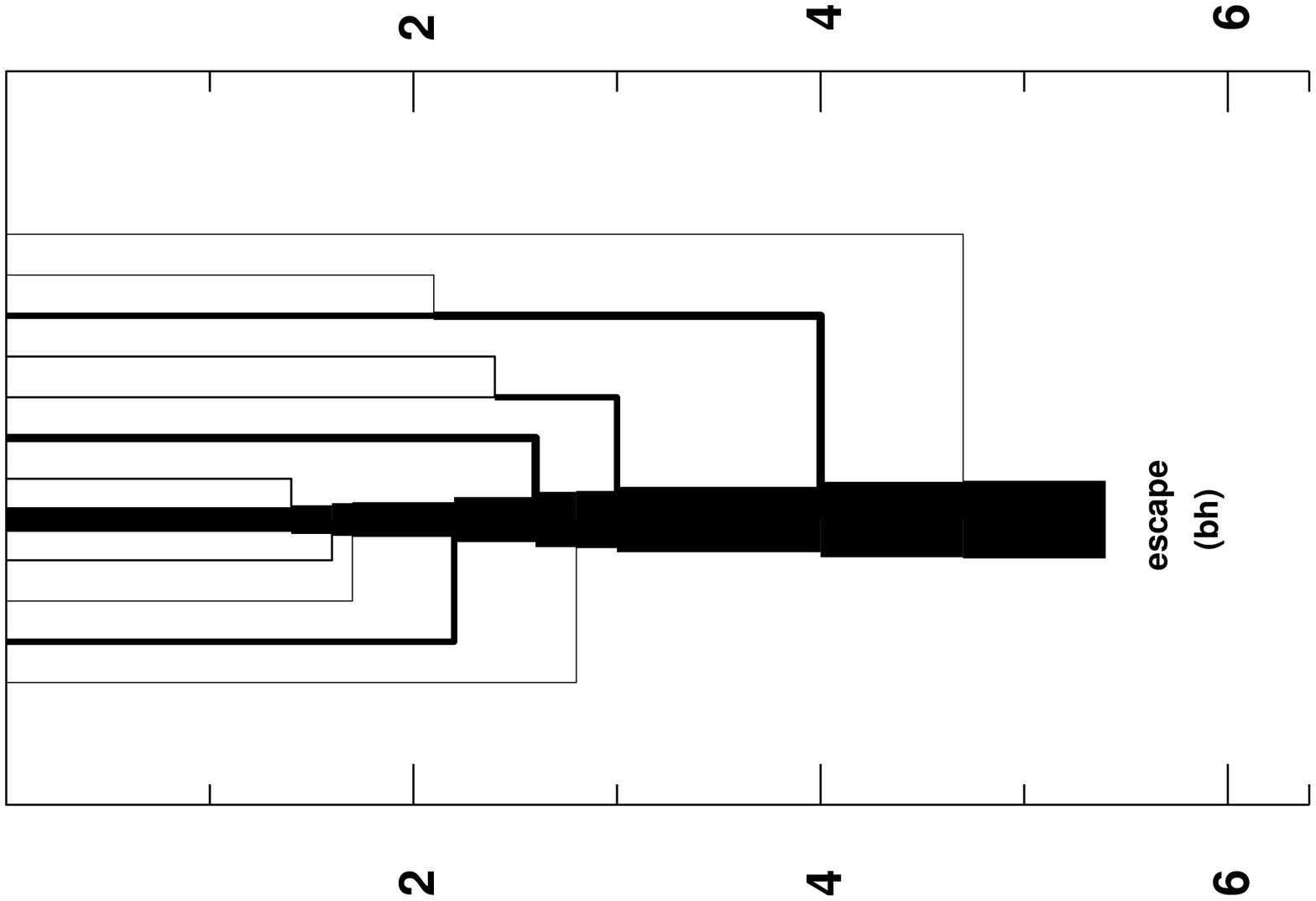,width=4.8cm,angle=-90}
\caption[]{Coagulation tree for models R90W4 (left) and 12k6A10
(right, see model Nk6X10 in Tab.\,\ref{Tab:models}).  Time is along
the vertical axis from zero age (top) to the moment the last collision
occurs (bottom). The thickness of the lines scales linear with
mass.  }
\label{Fig:runaway}
\end{figure}

Fig.\,\ref{Fig:runaway} presents the evolution tree for the collision
sequence for two of the models. In this (an all other) cases a single
object keeps colliding with other stars at a high rate.  Only a few
collisions occur between stars which do not finally coalesce with this
collisions runaway.  Only after the previous collision runaway is
ejected from the cluster, another can take its place (see left panel
in fig.\,\ref{Fig:runaway}).

%%%%%%%%%%%%%%%%%%%%
\section{Discussion}
In star clusters with a relaxation time smaller than a few 10\,Myr
collisions are quite common.  The majority of collisions involve one
selected object; the runaway merger.  This seed object involves
generally one of the initially most massive stars in the initial star
cluster. If the cluster is embedded in the tidal field of the Galaxy,
this selected star may eventually escape as a main sequence star or as
the leftover from a supernova (a neutron star or black hole). Once the
runaway merger has escaped another star will take its place until the
cluster dissolves in the tidal field of the Galaxy. In an isolated
cluster (or if the tidal fields on the cluster are small) it is harder
to get rid of the collision runaway and in these cases the runaway
merger may continue to grow in mass (see right panel in
Fig.\,\ref{Fig:runaway}).

Most of our calculations, however, were computed with rather strong
tidal forces, and these models dissolve within a few initial half mass
relaxation times.

\subsection{Observable characteristics of a collision runaway}

\subsection{Super blue stragglers}
The stars which have experienced several collision may be brighter and
bluer than any other star in the cluster, though detailed
characteristics are hard to sketch.  A few trivial characteristics may
be those of a blue straggler much more than twice the mass of the
clusters' turn off (Sills \& Lombardi
1997).\nocite{1997ApJ...484L..51S} For compact young star cluster such
a characteristic, however, is not very useful as the turn off is ill
defined for these systems.

The three most massive stars in R\,136 have spectral type WN4.5 and
appear to be younger than the other stars.  Their age is estimated to
be about 1\,Myr (Massey \& Hunter 1998; de Koter et al.\
1997).\nocite{1997ApJ...477..792D} These stars show violet absorption
edges, which are common for late type (WN8 and later) stars but highly
unusual for these early types (Conti et al.\
1983).\nocite{1983ApJ...268..228C} Also striking is that these stars
are unusually hydrogen rich (Massey \& Hunter 1998).  They are about
an order of magnitude brighter than normal for such stars.  Estimates
for their masses range from 112 to 155\,\msun\ (Chlebowski \& Garmany
1991; Vacca et al.\ 1996).
\nocite{1991ApJ...368..241C}\nocite{1996ApJ...460..914V} Two of them
lie well inside the core of the cluster; the third is at a projected
distance of about 0.6\,pc from the core.

Also the Quintuple cluster contains a very over-massive star, the
Pistol star, with and estimated mass of about 150\,\msun\, (Lang et
al. 1999).\nocite{1999AJ....118.2327L} Also this star appears to be
rather odd and somewhat younger than the rest of the cluster stars.

We aregue that both stars may be the result of (multiple) collisions,
causing a supermassive star which is slightly younger due to the
rejuvenation in the collision process.

\subsection{Super novae} 
Mass segregation causes massive cluster members to be preferentially
in the cluster center. This causes them to be excellent candidates for
subsequent collisions. In our calculations the mean mass of a
collision member is 10\,\msun. Stars with masses $\apgt 8$\,\msun\,
are expected to end their fuel burning life in a supernova, leaving a
neutron star or a black hole.  Collisions tend to reduce the number of
type Ib/c and type II supernovae.  Stars which normally would explode
in a supernova (zero age mass $\apgt 8$\,\msun) are likely collision
counterparts. A collision between two of such stars reduces therefore
the number of supernova.

\subsection{Hypernovae?}
In compact young star clusters the mean accumulating mass (mean
addition in mass per collision) is about 10\,\msun\, (see
Fig.\,\ref{Fig:model_rate}).  After \ncoll\, collisions the initially
most massive star (of 60\,\msun\, to 100\,\msun) in the stellar system
(the most likely collision runaway) has gained 10 \ncoll\,\msun\,
resulting in a total mass of $10\ncoll + (60-100)$\,\msun.  Each
collision may have rejuvenated the collision runaway somewhat, but
eventually its core will grow until it exceeds a Chandrasekhar mass in
iron, causing the star to explode in a supernova. The star itself is
likely to be rapidly rotation, possibly close to break-up (Sills \&
Lombardi 1997). The enormous mass of the exploding star together with
its high rotation rate may cause it to explode quite differently than
stars in isolation. Nakamura et al. (2000)\nocite{2000astro.ph.11184N}
argues that stars with an extremely high mass which are rotating very
rapidly may be excellent candidates for Hypernovae, possibly even
causing gamma-ray bursts.

\subsection{The formation of massive black holes}
As long as the collision runaway remains on the main sequence or one
of the giant branches its cross section for further collisions remains
large, due to the large size of the star. As soon as it explodes in a
supernova things become somewhat uncertain as we do not understand the
supernova mechanism very well.  It is generally accepted that such
massive stars form black holes but how much mass is lost from the
collapsing star is uncertain.  If the star loses a lot of mass in the
supernova it may eject itself from the cluster, because the star was
most likely to be a member of a close binary, which receives a high
runaway velocity upon the supernova explosion (see Blaauw
1961;\nocite{blaauw_1961} van den Heuvel et al.\, 2000; Portegies
Zwart 2000).\nocite{2000astro.ph.5245PZ}\nocite{2000astro.ph.5021PZ}
When little or no mass is lost in the supernova event the black hole
remains in the cluster and may continue to grow in mass via subsequent
collisions with other stars. Even though the size of the black hole
has decreased dramatically it will still dominate the collision rate as
its cross section is dominated by gravitational focusing.

If we allowed little mass loss upon the supernova
the formed black hole continues to grow in mass at about the same rate
as before the supernova, indicating that the collision
cross section is dominated by gravitational focusing.
Take into account that if no mass is lost upon the
supernova the black hole remains deep in the cluster core, where the
collision are occurring.

When more and more stars end their lifetime in a supernova the cluster
will lose more mass. This results in a local expansion of the cluster
core, finally terminating the collision efficiency. We therefore do
not expect that collisions remain dominating the cluster evolution for
more than a few 10 Myr. In older clusters binaries consisting of
previously formed black holes start to heat the cluster (see Portegies
Zwart \& McMillan 2000).\nocite{2000ApJ...528L..17P} This will all
finally result in a termination of the collision rate and causes the
cluster to expand.  Whether the cluster finally dissolves or remains
bound to experience another phase of core collapse depends on details
concerning the initial conditions of the star cluster. The further
evolution of the cluster and its central massive black hole is a
complicated matter and requires more study. We are quite happy that
this matter is beyond the scope of this paper.

\subsection{The effect of primordial binaries}
All our calculations started without primordial binaries, which, of
course, is an ill assumption. These clusters most likely contain a
rich population of binaries, which have been formed together with the
other stars in the cluster. For studying the collisional growth of a
central object primordial binaries might not be very important.  Of
course, the total mass in a binary is --on average-- 1.5 times higher
than the mean mass if single stars, but it will only make the increase
in mass of the runaway merger larger and the absence of primordial
binaries therefore reduces the effect of runaway growth.  On the other
hand, primordial binaries may heat the core at such a rate that
collision are prevented altogether. This is very unlikely, as by the
time binaries become hard enough to heat the cluster they are
extremely vulnerable to collisions. Also, our runaway merger was
generally a binary member, simply due to binary formation in the usual
3-body processes. We therefore expect the a rich population of
primordial binaries does not lower the observed collision rate in our
models. Detailed calculations in which we take a rich population of
primordial binaries into account are in progress.

\section{Conclusions}
We performed a large number of \nbody\, simulations which include the
effects of stellar evolution, binary evolution and the tidal field of
the Galaxy. The initial conditions are selected to represent the young
and compact stars clusters such as Arches and the Quintuplet systems.

Our model clusters are highly collisional, in the sense that
collisions occur at a very high rate, much higher than expected from
simple cross section arguments. The stars which are participating in
collisions are generally much more massive than the mean mass in the
cluster. Generally one of the most massive stars in the cluster
experiences multiple collisions until it either explodes in a
supernova or is ejected from the cluster by a dynamical encounter.  The
collision rate per star per million years in our models scales very
nicely with the initial relaxation time of the models, via:
\begin{equation}
	\rcoll = 2.2 \times 10^{-4} \trlx^{-1.0} \; \; 
		       [{\rm Myr}^{-1}].
\end{equation}
Here \trlx\, is the clusters' initial relaxation time in million
years. 

\acknowledgements 

SPZ is grateful to the Institute for Advanced Study, Drexel University
and Tokyo University for their hospitality and the use of their
GRAPE-4 hardware.  This work was supported by NASA through Hubble
Fellowship grant HF-01112.01-98A awarded by the Space Telescope
Science Institute, which is operated by the Association of
Universities for Research in Astronomy, by the Research for the Future
Program of Japan Society for the Promotion of Science
(JSPS-RFTP97P01102) and by NASA ATP grants NAG5-6964 and NAG5-9264.


\begin{thebibliography}{}

\bibitem[\protect\astroncite{Blaauw}{1961}]{blaauw_1961}
Blaauw, A. 1961, ban, 15, 265

\bibitem[\protect\astroncite{{Brandl} et~al.}{1996}]{1996ApJ...466..254B}
{Brandl}, B., {Sams}, B.~J., {Bertoldi}, F., {Eckart}, A., {Genzel}, R.,
  {Drapatz}, S., {Hofmann}, R., {Loewe}, M., {Quirrenbach}, A. 1996, \apj, 466,
  254

\bibitem[\protect\astroncite{{Campbell} et~al.}{1992}]{1992AJ....104.1721C}
{Campbell}, B., {Hunter}, D.~A., {Holtzman}, J.~A., {Lauer}, T.~R., {Shayer},
  E.~J., {Code}, A., {Faber}, S.~M., {Groth}, E.~J., {Light}, R.~M., {Lynds},
  R., {O'Neil}, E.~J., J., {Westphal}, J.~A. 1992, \aj, 104, 1721

\bibitem[\protect\astroncite{{Chlebowski} \&
  {Garmany}}{1991}]{1991ApJ...368..241C}
{Chlebowski}, T., {Garmany}, C.~D. 1991, \apj, 368, 241

\bibitem[\protect\astroncite{{Conti} et~al.}{1983}]{1983ApJ...268..228C}
{Conti}, P.~S., {Leep}, M.~E., {Perry}, D.~N. 1983, \apj, 268, 228

\bibitem[\protect\astroncite{{de Koter} et~al.}{1997}]{1997ApJ...477..792D}
{de Koter}, A., {Heap}, S.~R., {Hubeny}, I. 1997, \apj, 477, 792

\bibitem[\protect\astroncite{{Figer} et~al.}{1999a}]{1999ApJ...525..750F}
{Figer}, D.~F., {Kim}, S.~S., {Morris}, M., {Serabyn}, E., {Rich}, R.~M.,
  {McLean}, I.~S. 1999a, \apj, 525, 750

\bibitem[\protect\astroncite{{Figer} et~al.}{1999b}]{1999ApJ...514..202F}
{Figer}, D.~F., {McLean}, I.~S., {Morris}, M. 1999b, \apj, 514, 202

\bibitem[\protect\astroncite{{Giersz} \& {Heggie}}{1994}]{1994MNRAS.268..257G}
{Giersz}, M., {Heggie}, D.~C. 1994, \mnras, 268, 257

\bibitem[\protect\astroncite{{Giersz} \& {Heggie}}{1996}]{1996MNRAS.279.1037G}
{Giersz}, M., {Heggie}, D.~C. 1996, \mnras, 279, 1037

\bibitem[\protect\astroncite{{Glass} et~al.}{1987}]{1987MNRAS.227..373G}
{Glass}, I.~S., {Catchpole}, R.~M., {Whitelock}, P.~A. 1987, \mnras, 227, 373

\bibitem[\protect\astroncite{{Heggie} \& {Mathieu}}{1986}]{HM1986}
{Heggie}, D.~C., {Mathieu}, R. 1986, \mnras, in P. Hut, S. McMillan (eds.),
  Lecture Not. Phys 267, Springer-Verlag, Berlin

\bibitem[\protect\astroncite{{Heggie} \&
  {Ramamani}}{1995}]{1995MNRAS.272..317H}
{Heggie}, D.~C., {Ramamani}, N. 1995, \mnras, 272, 317

\bibitem[\protect\astroncite{{Hills}}{1975}]{1975AJ.....80..809H}
{Hills}, J.~G. 1975, \aj, 80, 809

\bibitem[\protect\astroncite{{Lang} et~al.}{1999}]{1999AJ....118.2327L}
{Lang}, C.~C., {Figer}, D.~F., {Goss}, W.~M., {Morris}, M. 1999, \aj, 118, 2327

\bibitem[\protect\astroncite{{Leonard}}{1989}]{1989AJ.....98..217L}
{Leonard}, P. J.~T. 1989, \aj, 98, 217

\bibitem[\protect\astroncite{{Lightman} \&
  {Shapiro}}{1977}]{1977ApJ...211..244L}
{Lightman}, A.~P., {Shapiro}, S.~L. 1977, \apj, 211, 244

\bibitem[\protect\astroncite{{Makino} et~al.}{1997}]{1997ApJ...480..432M}
{Makino}, J., {Taiji}, M., {Ebisuzaki}, T., {Sugimoto}, D. 1997, \apj, 480, 432

\bibitem[\protect\astroncite{{Massey} \& {Hunter}}{1998}]{1998ApJ...493..180M}
{Massey}, P., {Hunter}, D.~A. 1998, \apj, 493, 180

\bibitem[\protect\astroncite{{Nakamura} et~al.}{2000}]{2000astro.ph.11184N}
{Nakamura}, T., {Umeda}, H., {Iwamoto}, K., {Nomoto}, K., {Hashimoto}, M.,
  {Hix}, W.~R., {Thielemann}, F. 2000,
\newblock in Submitted to the Astrophysical Journal (August 11, 2000).,  11184

\bibitem[\protect\astroncite{{Portegies Zwart}
  et~al.}{2000a}]{2000astro.ph..8490P}
{Portegies Zwart}, S., {Makino}, J., {McMillan}, S., {Hut}, P. 2000a,
\newblock ApJ Letters in press. (astro-ph/0008490)

\bibitem[\protect\astroncite{{Portegies Zwart}
  et~al.}{2000b}]{2000astro.ph.5248PZ}
{Portegies Zwart}, S., {McMillan}, S., {Hut}, P., {Makino}, J. 2000b,
\newblock MNRAS in press.,  (astro-ph/0005248)

\bibitem[\protect\astroncite{{Portegies Zwart}}{2000}]{2000astro.ph.5021PZ}
{Portegies Zwart}, S.~F. 2000,
\newblock ApJ in press.,   (astro-ph/0005202)

\bibitem[\protect\astroncite{{Portegies Zwart} et~al.}{1999}]{pzmmh99}
{Portegies Zwart}, S.~F., {Makino}, J., {McMillan}, S. L.~W., {Hut}, P. 1999,
  \aap, 348, 117

\bibitem[\protect\astroncite{{Portegies Zwart} \&
  {McMillan}}{2000}]{2000ApJ...528L..17P}
{Portegies Zwart}, S.~F., {McMillan}, S. L.~W. 2000, \apjl, 528, L17

\bibitem[\protect\astroncite{Scalo}{1986}]{scalo86}
Scalo, J.~M. 1986, Fund. of Cosm. Phys., 11, 1

\bibitem[\protect\astroncite{{Sills} \& {Lombardi}}{1997}]{1997ApJ...484L..51S}
{Sills}, A., {Lombardi}, J.~C. 1997, \apjl, 484, L51

\bibitem[\protect\astroncite{{Vacca} et~al.}{1996}]{1996ApJ...460..914V}
{Vacca}, W.~D., {Garmany}, C.~D., {Shull}, J.~M. 1996, \apj, 460, 914

\bibitem[\protect\astroncite{{van Den Heuvel}
  et~al.}{2000}]{2000astro.ph.5245PZ}
{van Den Heuvel}, E., {Portegies Zwart}, S.~F., {Batacharya}, D., {Kaper}, L.
  2000,
\newblock A\&A in press  (astro-ph/0005245)

\end{thebibliography}
\end{document}